\documentclass[%
 reprint,
showpacs,
 amsmath,amssymb,
 aps,
]{revtex4-1}

\usepackage{graphicx}
\usepackage{dcolumn}
\usepackage{epstopdf}
\usepackage{bm}

\begin{document}

\preprint{APS/123-QED}

\title{Joint effect of polarization and the propagation path of a light beam on its intrinsic structure}

\author{Sarkew Abdulkareem}
\affiliation{%
 Department of Optics and Spectroscopy, \\ South Ural State University,  76 Lenin Av., Chelyabinsk, 454080 Russia
}%


\author{Nataliya Kundikova}
 \email{kundikovand@susu.ru}
\affiliation{
 Nonlinear Optics Laboratory, Institute of Electrophysics, Ural Branch of the Russian Academy of Sciences, 106 Amundsena Str., Ekaterinburg, 620016 Russia
}%
\affiliation{
 Department of Optics and Spectroscopy, \\ South Ural State University, 76 Lenin Av., Chelyabinsk, 454080 Russia
}%
%

\date{\today}

\begin{abstract}
The well-known effects of the spin-orbit interaction of light are manifestations of pair mutual influence of the three types of the angular momentum of light, namely, the spin angular momentum, the extrinsic  orbital angular momentum and the intrinsic orbital angular momentum. 
Here we propose the convenient classification of the effects of the spin-orbit interaction of light and we observe one of the new effects in the frame of this classification, which is determined by the joint  influence of two types of the angular momentum on the third type of the angular momentum, namely, the influence of the spin angular momentum and the extrinsic orbital angular momentum on the  intrinsic orbital angular momentum.
We experimentally studied the propagation of circularly polarized light through an optical fiber coiled into a helix. We have found that the spin angular momentum and the helix  parameters affect the spatial structure of the radiation  transmitted through the optical fiber.
We found out that the structure of the light field rotates when changing the sign of circular polarization. The angle of rotation depends on the parameters of the helix.
The results can be used to develop the  general theory of spinning particles and can find application in metrology  methods and nanooptics devices.

\end{abstract}

\maketitle



Structured light beams carry three types of angular
 momentum  \cite{Beth1936,Allen2003,Bliokh2006,Bekshaev2011}. 
The spin angular momentum is associated with polarization, the extrinsic  orbital angular momentum is determined by the propagation path of the light beam, and the intrinsic orbital angular momentum is determined by the structure of the  light field of the beam \cite{Bliokh2006}. 
The effect of one of the angular momenta on another angular momentum  leads to the spin-orbit interaction of light (a photon) \cite{ Dugin1991,Liberman1992}. 
There are six variants of such effects.

The {\it spin angular momentum} affects  the extrinsic orbital angular momentum, the effect  can be observed as the longitudinal shift of the centroid of  a linearly polarized light beam  and the transverse shift of  the centroid of a circularly polarized light beam  at reflection and refraction and in focused light beams.
These shifts are known as the Goos-Hanchen shift \cite{ Goos1947}, the Imbert-Fedorov shift \cite{ Fedorov55,Kristoffel1956,Imbert1970}, the Hall effect for light \cite{Onoda2004}, the optical Magnus effect \cite{ Dooghin1992} and the shift of the beam waist   \cite{BaranovaN.B.1994,ZeldovichB.Ya.1994,Kundikova1995}.

The {\it extrinsic orbital angular momentum} affects the spin angular momentum, the effect manifests itself as the rotation of the linear polarization of light when changing the light propagation path \cite{Rytov1938,Vladimirskii1941,Chiao1986,Berry1987}.
The effect is known as the Rytov-Vladimirski-Berry-Chao-Wu-Tomita geometric polarization rotation. It can be observed in a  single mode fiber, coiled into a helix \cite{Tomita1986}, or in a multimode optical fiber \cite{1063-7818-25-2-A24}.

The {\it intrinsic orbital angular momentum } affects the extrinsic orbital angular momentum, the effect manifests itself as the shift of the centroid of  a vortex light beam under reflection and refraction \cite{Fedoseyev2001,Bliokh2006,Merano2010,Dennis2012}.

The {\it extrinsic orbital angular momentum} affects the intrinsic orbital angular momentum, the effect manifests itself as the change of the  beam field structure   when changing the  propagation path of a beam \cite{Bliokh2006,Alekseyev2007,Kataevskaya1995,Bolshakov2011}.
The rotation of the speckle pattern of the light transmitted through the optical fiber, coiled into a helix, was experimentally observed when changing the pitch of the helix \cite{Kataevskaya1995,Bolshakov2011}. 

Interaction of the {\it spin angular momentum} with the intrinsic orbital angular momentum manifests itself as the transformation of the circular polarized beam 
of zero vorticity into the linearly polarized beam of non-zero vorticity
\cite{Darsht95a,Marrucci2006,Zhao2007,Vuong2010,Bliokh:11,Kobayashi2012,Vasylkiv2013}. 
Such transformation can be observed in anisotropic inhomogeneous medium \cite{Marrucci2006,Vasylkiv2013}, in fibers \cite{ Darsht95a}, in focused beams \cite{Zhao2007,Bliokh:11}  and under light scattering \cite{Vuong2010,Bliokh:11}.

As for inverse effect, the transformation of the {\it intrinsic orbital angular momentum} into the spin angular momentum was observed with vector autofocusing Airy beams \cite{Liu2013}. 

Recently published review \cite{Bliokh2015}  provides considerably more detailed information on the spin-orbit interaction of light.

The study of the spin-orbit interaction of light is of  great interest
because experimental observations in the optical range are much easier, and the results can be used to develop the theory of spinning  particles and for the search of new effects \cite{Berard2006,Duval2007,Andrzejewski2015,Duval2015}.

The effects of the spin orbit interaction of light are sufficiently small and neglected in terms of geometrical optics. However, when operating at subwavelength scales, these effects should be taken into account. They are very sensitive to a change in the physical state of systems and are promising for application in high-precision metrology. 

They can be used to determine the spatial distribution of electronic spin states in semiconductors \cite{ Menard2009}, to determine the parameters of films \cite{ Zhou2012,Bolshakov2015}, to image graphene layers \cite{ Zhou2012a}, and to investigate topological insulators \cite{PhysRevA.88.053840}. The effects should be taken into account when designing nanophotonics devices and can be used for such devices creation \cite{Corrielli2014}.

Here we report the results of an experimental study of the joint effect of two parts of the angular momentum on the third, namely, the joint effect of the spin angular momentum and the extrinsic orbital angular momentum on the intrinsic orbital angular momentum.

We examined the optical Magnus effect \cite{ Dooghin1992}  in the optical fiber, coiled into a helix, and have found out the effect of polarization (spin angular momentum) and the helix parameters (extrinsic orbital angular momentum) on the structure of the light field (intrinsic orbital angular momentum), transmitted through the optical fiber.
To increase the accuracy of the measurements, we used a method based on the wavefront conjugation \cite{Darscht2003,Asselborn2006a}.
We have found that the optical Magnus effect decreases in a negative helix and  increases in a positive helix.

The observed effect is one of three possible effects of the joint impact of two types of the angular momentum on the third type of the angular momentum.
Our classification of the effects of the spin-orbit interaction of light shows that it is possible to find out two new effects of the spin-orbit interaction of light. 
These are 1) the joint influence of the spin angular momentum and the intrinsic orbital angular  momentum  on the extrinsic orbital angular momentum and 2) the joint influence of the extrinsic orbital angular momentum and the intrinsic  orbital angular momentum on the spin angular momentum.

Optical Magnus effect \cite{Dugin1991,Dooghin1992,Liberman1992},
 which manifests itself as the rotation of the speckle pattern of circularly polarized light transmitted through a multimode optical fiber under the  change of  the sign of the circular polarization, is the result of the accumulation of transverse spatial shifts under the circularly polarized light propagation through an optical fiber. A multimode optical fiber can be easily coiled into a helix; as a result, the topological optical activity arises due to the Berry phase.
  The different refractive indices for the right and left circular polarized light should influence the polarized light propagation through the fiber \cite{Tomita1986, Chiao1986,Alekseyev2007}.

Let us consider the polarized light propagation in a multimode optical fiber with a step index profile.
 In such a fiber, the light field inside the fiber is a superposition of modes ${J_{\left| l \right|}}\left( r \right)\exp \left( {il\varphi } \right)$, where $r,\varphi $ are the polar coordinates, ${J_{\left| l \right|}}\left( r \right)$ is the Bessel function, $l$ is a topological charge or an orbital angular momentum, $ - {l_{\max }} \le l \le {l_{\max }}$, ${l_{\max }} = \left( {{{2\pi \rho } \mathord{\left/ {\vphantom {{2\pi \rho } \lambda }} \right.  \kern-\nulldelimiterspace} \lambda }} \right)\sqrt {2{n_{{\rm{co}}}}\delta n} $, $\rho$ is the radius of the fiber core , $\lambda $ is a wavelength, $\delta n = {n_{{\rm{co}}}} - {n_{{\rm{cl}}}}$ is the difference in the refractive indices of the core $n_{{\rm{co}}}$ and cladding $n_{{\rm{cl}}}$. One can neglect the modes with $l = 0, \pm 1$ in a multimode optical fiber, and then circularly polarized field ${{\bf{E}}^\sigma }\left( {r,\varphi ,z} \right)$ inside the fiber is kept constant and has the following form \cite{Snyder1983,Darsht95a}:
\begin{widetext}
\begin{equation}
{{\bf{E}}^\sigma }\left( {r,\varphi ,z} \right) = \frac{{{{\bf{e}}_x} + i\sigma {{\bf{e}}_y}}}{{\sqrt 2 }}\sum\limits_{l \ne 0, \pm 1} {\sum\limits_N {{C_{l,N}}} } {e^{il\varphi }}{J_{\left| l \right|}}\left( r \right) \times \exp \left[ {iz\left( {{\beta _{l,N}} + \delta \beta _{l,N}^\sigma } \right)} \right].
\label{eq1}
\end{equation}
\end{widetext}
Here $\sigma  =  + 1$ stands for the right circularly polarized light,  $\sigma  =  - 1$ stands for the left circularly polarized light, ${C_{l,N}}$ are  complex coefficients that determine the contribution of each mode in the light field, ${\beta _{l,N}}$ are the propagation constants of light in the fiber and $\delta \beta _{l,N}^\sigma $ are the polarization corrections to 
propagation constants ${\beta _{l,N}}$. Analytical expressions for ${\beta _{l,N}}$ and $\delta \beta _{l,N}^\sigma $ can be found in \cite{Dooghin1992, Snyder1983}.

If a multimode optical fiber is coiled into a helix with  diameter $d$ and
 pitch $h$, then  additional corrections to the propagation constants arise from the Berry phase. It is obvious, that the additional corrections depend on the sign of the circular polarization $\sigma $, and the sense of the helix $\gamma $. Let $\gamma  =  + 1$ stands for the right helix and $\gamma  =  - 1$ stands for the left helix. Then it is easy to show that the corrections to the propagation constants $\delta \beta _{\rm B}^{\sigma ,\gamma }$ caused by the Berry phase  have the following form:
\begin{equation}
\delta \beta _{\rm{B}}^{\sigma ,\gamma } = \sigma \gamma \frac{{2\pi h}}{{\left( {{\pi  ^2}{d^2} + {h^2}} \right)}}. 
\label{eq2}
\end{equation} 
Taking into  consideration the added correction $\delta \beta _{\rm{B}}^{\sigma ,\gamma }$ to the propagation constants, we obtain  Eq. (\ref{eq1}) as follows:
\begin{widetext}
\begin{equation}
{{\bf{E}}^\sigma }\left( {r,\varphi ,z} \right) = \frac{{{{\bf{e}}_x} + i\sigma {{\bf{e}}_y}}}{{\sqrt 2 }}\sum\limits_{l \ne 0, \pm 1} {\sum\limits_N {{C_{l,N}}} } {e^{il\varphi }}{F_{\left| l \right|,N}}\left( r \right) \times \exp \left[ {iz\left( {{\beta _{l,N}} + \delta \beta _{l,N}^\sigma  + \delta \beta _{\rm{B}}^{\sigma ,\gamma }} \right)} \right].
\end{equation}
\end{widetext}
Let us  analyze the magnitude of the corrections to the propagation constants for the fiber, which was used for the first experimental observation of the optical Magnus effect \cite{Dugin1991}. The fiber had the following parameters. The refractive index of the core ${n_{{\rm{co}}}}{\rm{ = 1}}{\rm{.500}}$, the refractive index of the cladding ${n_{{\rm{cl}}}}{\rm{ = 1}}{\rm{.494}}$, the fiber core radius $\rho  = 100\;{\rm{\mu m}}$, the wavelength  $\lambda  = 633$ nm. The propagation constants  values belong to the range determined by the refractive indices of the core and cladding:
\begin{equation}
{n_{{\rm{cl}}}}\frac{{2\pi }}{\lambda } \le {\beta _{lN}} \le {n_{{\rm{co}}}}\frac{{2\pi }}{\lambda },
\end{equation}
or
\begin{equation}
1.4822 \times {10^5}{\rm{c}}{{\rm{m}}^{ - 1}} \le {\beta _{lN}} \le 1.4882 \times {10^5}{\rm{c}}{{\rm{m}}^{ - 1}}.
\end{equation}

In accordance with the calculations carried out in Ref. \cite{Dooghin1992} the absolute values of $\delta \beta _{l,N}^\sigma $ are in the range of  $0 <  < \left| {\delta \beta _{l,N}^\sigma } \right| <  < 0.070$  ${\rm cm^{-1}}$.  According to Eq.(\ref{eq2}), the absolute values of  $\delta \beta _{\rm{B}}^{\sigma ,\gamma }$ are in the range of  $0.0 <  < \left| {\delta \beta _{\rm{B}}^{\sigma ,\gamma }} \right| <  < 0.058$ ${\rm cm^{-1}}$ when the radius of the helix is equal to 5 cm, and the helix pitch varies from 0 to 10 cm. These helix parameters were  used for the first experimental observation of the speckle-pattern rotation in the fiber coiled into a helix \cite{Kataevskaya1995}.

To carry out experimental investigation we used a fiber with the following parameters:
fiber core radius $\rho  = 100$ $ \rm \mu m$, core refractive index  ${n_{{\rm{co}}}} = 1.458$,  cladding refractive index  ${n_{{\rm{cl}}}} = 1.441$,  wavelength $\lambda  = 532$ nm.

In order to determine the angle of the speckle-pattern rotation with high accuracy, we used the method based on the optical phase conjugation of the radiation transmitted through an optical fiber \cite{Darscht2003,Asselborn2006a}. The phase conjugation of circular polarized light transmitted through a multimode optical fiber allows to invert light propagation and obtain a narrow light beam at the other fiber end. As a result, the optical Magnus effect leads to the rotation of only one spot around the fiber axis under the circular polarization sign changing. This method makes it possible to work with only one spot instead of the whole speckle-pattern and to observe relatively small changes in the behavior of the speckle-pattern.

\begin{figure}[h]
\includegraphics{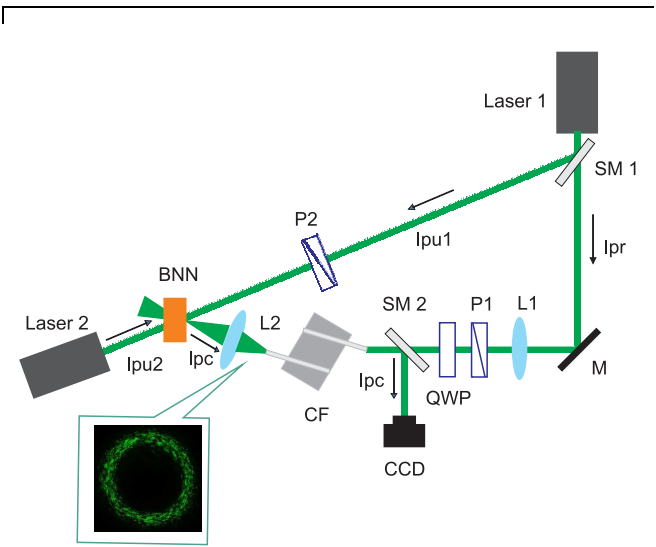}
\caption{\label{fig:epsart3} 
Experimental setup. SM, semitransparent  mirrors; M, mirror; L, lenses; BNN,
  photorefractive crystal ${\rm{B}}{{\rm{a}}_{\rm{2}}}{\rm{NaN}}{{\rm{b}}_{\rm{5}}}{{\rm{O}}_{{\rm{15}}}}$; P, polarizer; CF, optical fiber, coiled into a helix; 
  QWP, adjustable quarter-wave plate;   CCD, CCD matrix. The inset depict the speckle pattern of light transmitted through a coiled optical fiber.}
\end{figure}

The experimental  setup for the investigation of the optical Magnus effect in a coiled fiber  is shown in Fig. \ref{fig:epsart3}. 
Nd:YAG laser radiation at 
second harmonic wavelength $\lambda  = 532$ nm was used. It was convenient to use two Nd:YAG lasers. The radiation of the first laser passes through  semi-transparent mirror SM1 and is divided into two beams. The transmitted part of the beam is sent to the polarizing system consisting of  polarizer P1 and  adjustable quarter-wave plate QWP
 \cite{Bibikova2013}, which is then used as probe beam $I_{\rm pr}$. Circularly polarized probe beam $I_{ \rm pr}$ is focused by  lens L1 at the input end of the fiber   at angle $\vartheta  = {9.7^ \circ }$  to the fiber axis.

The fiber was coiled into a uniform helix by winding onto a cylinder of a fixed diameter.  The cylinder diameter  was equal to $d = 10$ cm. In order to form a closed path in momentum space, the propagation directions of the input and output ends of the fiber were kept identical.  Solid angle $\Omega $ subtended by the tangential vector to the curved trajectory at the unit sphere in the momentum space  was determined  in a way described in Ref. \cite{Tomita1986}. Angle $\Omega $ can be changed by changing the helix parameters.

The output speckle pattern is focused by lens L2 at the front  face of  photorefractive crystal ${\rm{B}}{{\rm{a}}_{\rm{2}}}{\rm{NaN}}{{\rm{b}}_{\rm{5}}}{{\rm{O}}_{{\rm{15}}}}$ (BNN). The reflected part of the radiation of the first laser, being passed through  polarize P2,  is used as   pump beam $I_{\rm pu}1$. Pump beam $I_{\rm pu}1$, linearly polarized in the horizontal plane, impinges on the front face of  photorefractive crystal BNN. The angle  
between  probe $I_{\rm pr}$ and pump $I_{\rm pu}1$ beams  is equal to ${21^ \circ }$. The linearly polarized part of  probe beam $I_{\rm pr}$ and  linearly polarized pump beam $I_{\rm pu}1$ record a hologram in  photorefractive crystal BNN.

The recorded hologram is illuminated by the counter propagating second pump beam $I_{\rm pu}2$ of the second laser. This beam is linearly polarized in the horizontal plane. As a result of beam $I_{\rm pu}2$ diffraction on the recorded hologram, conjugated beam $I_{\rm pc}$ propagates through the fiber in the opposite direction. 

\begin{figure}[h]
\includegraphics{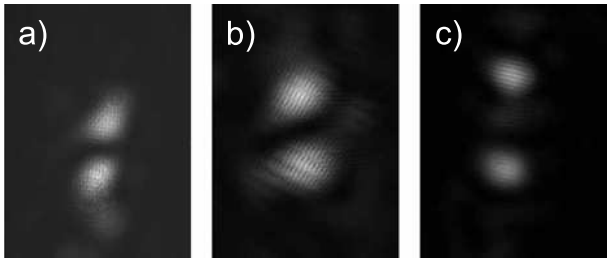}
\caption{\label{fig:epsart5} 
Images of the conjugated wave registered by a CCD camera. The fiber length was 65 cm, the right helix diameter was 10 cm, the  helix pitch (solid angle $\Omega $) was (a) 2 cm (0.4 sr), (b)  4 cm (0.79 sr) and (c) 6 cm (1.18 sr). Angle $\vartheta$ of light incidence at the fiber input was equal to ${9.7^ \circ }$.}
\end{figure}

The linearly polarized radiation is the superposition of two circularly polarized beams of equal intensity and different signs of circular polarization. 
Due to the optical Magnus effect, the circularly polarized light of the opposite circulation signs propagates along different trajectories and two beams of equal intensity and the opposite sign of the circular polarization can be seen at the fiber exit instead of only one linearly polarized beam. Images of the beams recorded by CCD camera after reflection from a semitransparent mirror SM2 are shown in Fig. \ref{fig:epsart5}.
 Images were obtained for  the fiber coiled  into a right helix of one coil. The helix diameter was 10 cm, the helix pitch was  2, 4  and 6 cm, the angle of incidence at the fiber end $\vartheta  = {9.7^ \circ }$,
 the fiber  length being 65 cm.

Figure \ref{fig:epsart5} shows that the distance between two beams increases along with the increase of the helix pitch, or  solid angle $\Omega $, subtended by one helix coil. To determine the angle of the speckle pattern rotation, we measured distance between the observed beams centroid  and the distance between the fiber end and the CCD camera \cite{Darscht2003,Asselborn2006a}. In Figure \ref{fig:epsart5}, the distances between beams correspond to the angles of rotation $\varphi  = {\rm{3}}{\rm{.6}}{{\rm{1}}^ \circ }$, $\varphi  = {\rm{3}}{\rm{.9}}{{\rm{4}}^ \circ }$ and $\varphi  = {\rm{4}}{\rm{.3}}{{\rm{0}}^ \circ }$,  for the helix pitches $h$ ($\Omega$) 
 of 2 cm (0.4 sr), 4 cm (0.79 sr) and 6 cm (1.18 sr), respectively. As it can be seen in Figure \ref{fig:epsart5}, the used method provides  a  highly accurate determination of angle $\varphi $. 
The change of rotation angle $\varphi$  by angle $\Delta \varphi  = {0.69^ \circ }$ (Fig. \ref{fig:epsart5}a and  \ref{fig:epsart5}c) results in the two-fold increase of the distance between the beam images. 

To determine the polarization state of each of the beams, the polarization system was installed in front of the CCD camera. The polarization system consisted of the adjustable quarter-wave plate and the polarizer  was used as a circular analyzer which selects either the left or right circularly polarized radiation. The beams turned out to have orthogonal circular polarization. In Figure \ref{fig:epsart5}, the upper beams have the right  circular polarization, whereas the lower beams have the left circular polarization.

\begin{figure}[h]
\includegraphics{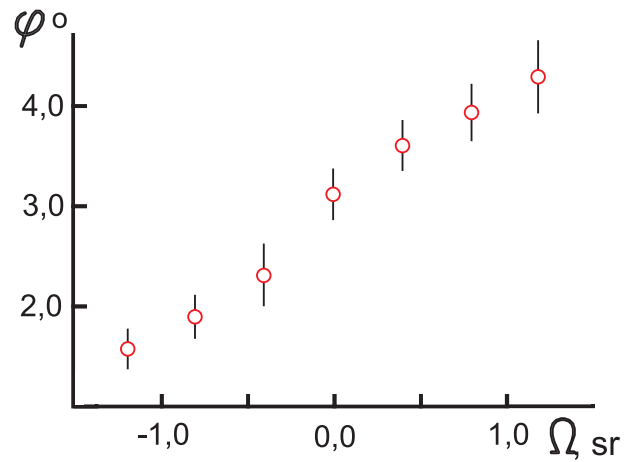}
\caption{\label{fig:epsart6} 
The dependence of rotation angle $\varphi $  of the speckle pattern of light transmitted through the optical fiber, coiled into a helix, under the sign of the circular polarization change on  solid angle $\Omega $ subtended by one helix coil in the momentum space. The diameter of one coil of the  uniform right and left helix $d = 10$ cm, fiber length being 65 cm.}
\end{figure}

The similar experiments were carried out for the fiber, coiled into the left helix. Figure \ref{fig:epsart6}
shows the dependence of  rotation angle $\varphi $ of  the speckle pattern  on 
solid angle $\Omega $. Positive values of solid angle $\Omega $ correspond to the right helix and negative values of solid angle $\Omega $ correspond to the left helix. 
Rotation angle  $\varphi $ at  point $\Omega  = 0$ coincides with rotation angle $\varphi $ in the rectilinear fiber. Figure \ref{fig:epsart6} shows that 
angle $\varphi $ increases along with the increase 
of solid angle  module $\left| \Omega  \right|$  for  the right helix, whereas  angle $\varphi $ decreases along with the increase of solid angle  module  $\left| \Omega  \right|$ for  the left helix. Figure \ref{fig:epsart6} shows that the optical Magnus effect depends on the propagation path and the helix sign, it linearly depends on the  helix pitch, decreases in a negative helix and  increases in a positive helix.
 
 Our experimental study of the optical Magnus effect  in the optical fiber, coiled into a helix, clearly  demonstrates  the joint effect of polarization (spin angular momentum) and the helix parameters (extrinsic orbital angular momentum) on the structure of the light field (intrinsic orbital angular momentum), transmitted through the optical fiber.

In conclusion, 
we classified all effects of the spin-orbit interaction of light and pointed out that 
three new effects can be found. These effects are  the joint  influence of two types of angular momentum on the third type of the angular momentum, namely, the joint influence of the spin angular momentum and extrinsic orbital angular momentum on the  intrinsic orbital angular momentum;
 the joint influence of the spin angular momentum and the intrinsic orbital angular  momentum  on the extrinsic orbital angular momentum;   
the  joint influence of the extrinsic orbital angular momentum and the intrinsic  orbital angular momentum on the spin angular momentum.

We experimentally observed  one of these effects, determined by the joint  influence 
of the spin angular momentum and extrinsic orbital angular momentum on the  intrinsic orbital angular momentum. 
We have studied the optical Magnus effect in a  fiber, coiled into a helix. 
We have found that the optical Magnus effect in a coiled fiber depends on the propagation path and the helix sign. It linearly depends on the  helix pitch, decreases in a negative helix and  increases in a positive helix.

 The authors are grateful to Dr. Sergei Asselborn and Kristina Mikhailyuk for their help in the experiment. We also express our appreciation  to Dr. Victor Kireev for valuable discussions. 

This work was partly carried out within the scope of the topic of State Assignment No. 0389-2014-0030.

%


\end{document}